\documentclass[sigconf]{acmart}
\usepackage{soul}
\usepackage{amsmath}
\usepackage{amsthm}
\usepackage{url}
\usepackage{hyperref}
\usepackage{inputenc}
\usepackage{caption}
\usepackage{graphicx}
\usepackage{booktabs}
\usepackage{algorithm}
\usepackage{algorithmic}
\usepackage[switch]{lineno}
\usepackage{multirow}
\usepackage{multicol}
\usepackage{enumitem}
\usepackage{subcaption}
\usepackage{threeparttable}
\usepackage{xspace}
\usepackage{appendix}
\usepackage{color}
\usepackage[normalem]{ulem}

\renewcommand\footnotetextcopyrightpermission[1]{}

\AtBeginDocument{%
  }

\begin{document}
\title{BiVRec: Bidirectional View-based Multimodal Sequential Recommendation}

\author{Jiaxi Hu}
\email{jiaxihu2-c@my.cityu.edu.hk}
\affiliation{%
  \institution{City University of Hong Kong}
  \country{}
  }

\author{Jingtong Gao}
\email{jt.g@my.cityu.edu.hk}
\affiliation{%
  \institution{City University of Hong Kong}
  \country{}
}

\author{Xiangyu Zhao}
\authornote{Corresponding Author.}
\email{xianzhao@cityu.edu.hk}
\affiliation{%
  \institution{City University of Hong Kong}
  \country{}
}

\author{Yuehong Hu}
\email{yhu322@connect.hkust-gz.edu.cn}
\affiliation{%
  \institution{Hong Kong University of Science and Technology (Guangzhou)}
  \country{}
  }

\author{Yuxuan Liang}
\email{yuxuanliang@outlook.com}
\affiliation{%
  \institution{Hong Kong University of Science and Technology (Guangzhou)}
  \country{}
  }

\author{Yiqi Wang}
\email{wangy206@msu.edu}
\affiliation{%
  \institution{Michigan State University}
  \country{}
  }

\author{Ming He}
\email{Shanghai Jiaotong University}
\affiliation{%
  \institution{}
  \country{}
  }

\author{Zitao Liu}
\email{liuzitao@jnu.edu.cn}
\affiliation{%
  \institution{Jinan University}
  \country{}
  }

\author{Hongzhi Yin}
\email{h.yin1@uq.edu.au}
\affiliation{%
  \institution{The University of Queensland}
  \country{}
  }
\newcommand{\etal}{\emph{et al.}\xspace}
\newcommand{\eg}{\emph{e.g.,}\xspace}
\newcommand{\ie}{\emph{i.e.,}\xspace}
\newcommand{\etc}{\emph{etc.}\xspace}

\renewcommand{\shortauthors}{Trovato et al.}

\begin{abstract}
The integration of multimodal information into sequential recommender systems has attracted significant attention in recent research. 
In the initial stages of multimodal sequential recommendation models, the mainstream paradigm was ID-dominant recommendations, wherein multimodal information was fused as side information.
However, due to their limitations in terms of transferability and information intrusion, another paradigm emerged, wherein multimodal features were employed directly for recommendation, enabling recommendation across datasets. 
Nonetheless, it overlooked user ID information, resulting in low information utilization and high training costs. 
To this end, we propose an innovative framework, BivRec, that jointly trains the recommendation tasks in both ID and multimodal views, leveraging their synergistic relationship to enhance recommendation performance bidirectionally. 
To tackle the information heterogeneity issue, we first construct structured user interest representations and then learn the synergistic relationship between them. 
Specifically, BivRec comprises three modules: 
Multi-scale Interest Embedding, comprehensively modeling user interests by expanding user interaction sequences with multi-scale patching;
Intra-View Interest Decomposition, constructing highly structured interest representations using carefully designed Gaussian attention and Cluster attention; 
and Cross-View Interest Learning, learning the synergistic relationship between the two recommendation views through coarse-grained overall semantic similarity and fine-grained interest allocation similarity
BiVRec achieves state-of-the-art performance on five datasets and showcases various practical advantages. 

\end{abstract}


\maketitle
\section{Introduction}
Recommender systems have demonstrated their effectiveness in mitigating the issue of information overload across diverse data-driven platforms \cite{liu2022multi}. Among these systems, sequential recommender systems play a pivotal role \cite{kang2018self}, which capitalize on users' historical interaction sequences to effectively model their preferences and interests. Notably, there has been a burgeoning interest in enhancing the efficacy of sequential recommendation models through the integration of multimodal information, such as visual or textual descriptions of items. 

\begin{figure}[t]
    \centering
    \includegraphics[width=0.95\linewidth]{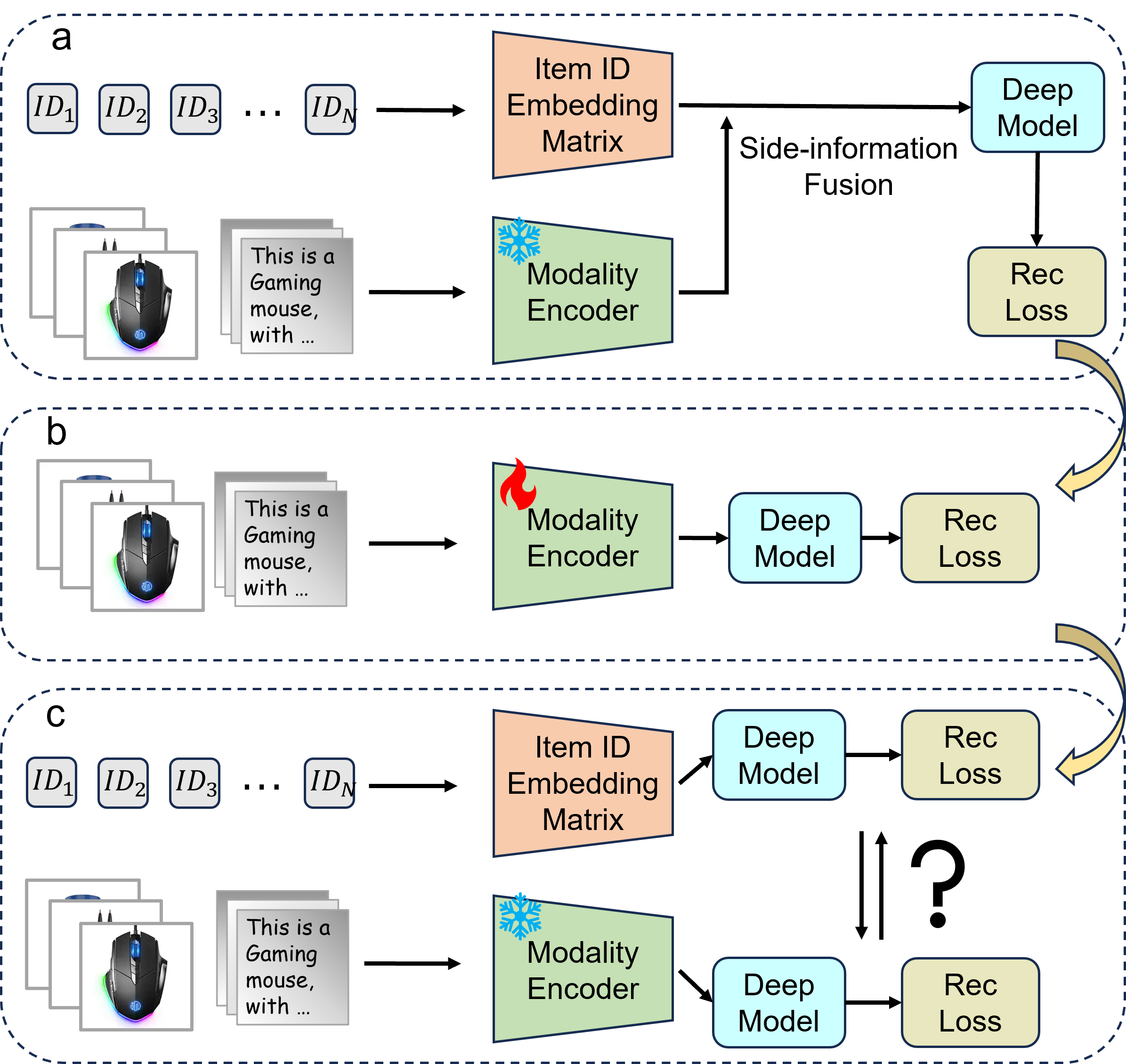}
    \vspace{-3mm}
    \caption{Exsiting (a) \& (b) and our proposed (c) paradigms for MSR. The symbols of snowflake and flame indicate whether the parameters are frozen or not.} 
    \label{fig:ID_MM}
    \vspace{-3mm}
\end{figure}

Early landscape of multimodal sequential recommendation (MSR) models \cite{liu2021noninvasive, rashed2022carca, pan2022multimodal,liu2023multimodal} predominantly rely on item-ID\footnote{In this paper, we will use ``ID'' for simplicity} information, due to the dominant position of ID-based recommendation paradigms.
As depicted in Figure \ref{fig:ID_MM} (a), this paradigm treats multimodal features as side information, with the goal of seamlessly integrating pre-trained multimodal representations into the ID information through concatenation \cite{rashed2022carca}, summation \cite{chen2022intent}, and attention mechanisms \cite{liu2021noninvasive}, etc., to provide more precise modeling of users' interests. 
However, this approach has notable limitations:
(i) ID information lacks the characteristics of cross-dataset existence, making it incompatible with the prevailing trend of universal large models.
(ii) Increasingly, research efforts have raised concerns about the direct fusion of multimodal and ID information \cite{liu2021noninvasive, xie2022decoupled, yuan2023go}, highlighting its potential to disrupt the underlying information structure.

In response to above limitations, another recommendation paradigm has emerged, stemming from developments in the fields of computer vision (CV) and natural language processing (NLP). Illustrated in Figure \ref{fig:ID_MM} (b), this framework leans towards pure multimodal recommendation tasks \cite{hou2022towards, wang2022transrec, yuan2023go, zhang2023multimodal, hou2023learning, song2023self}, wherein item multimodal features are directly employed as item representations. However, this framework is not without its challenges:
(i) While pure multimodal-based models effectively address the issue of information transferability, they completely disregard ID information, which is the most prevalent type of information, resulting in suboptimal information utilization.
(ii) Such models often rely on potent multimodal backbone encoders and typically necessitate end-to-end fine-tuning to achieve performance levels comparable to ID-based recommendation models \cite{xie2022contrastive, wang2022transrec, yuan2023go}. Consequently, their training costs can be prohibitively high in practical applications.

Upon revisiting the two aforementioned paradigms, a natural question arises: \textit{Can we amalgamate the merits of both paradigms while simultaneously addressing their inherent limitations to formulate a novel comprehensive paradigm?} This paper provides a definitive answer to this question by introducing a novel approach: the joint training of two recommendation paradigms (views) within the realms of both ID and multimodal perspectives, as illustrated in Figure \ref{fig:ID_MM} (c). This innovative framework effectively mitigates the shortcomings associated with the preceding two paradigms:
(i) The multimodal recommendation branch inherits the cross-dataset recommendation capability inherent in multimodal information, allowing for seamless integration with widely adopted multimodal backbone models.
(ii) Instead of directly fusing the two types of information, this framework harnesses the synergistic relationship between the two recommendation views to enhance the performance of both tasks bidirectionally.
(iii) This framework offers flexibility in selecting the information type employed for downstream recommendation task, thereby ensuring adaptability to diverse recommendation scenarios.
(iv) In contrast to the second aforementioned approach, which entails fine-tuning the parameters of the multimodal encoder, this framework maintains the modality encoders frozen, with the primary focus on modeling user intent through subsequent deep recommendation models. This approach significantly reduces training costs, given that the number of parameters in deep recommendation models is substantially smaller than that in the multimodal encoder.

The primary challenge encountered when jointly training recommendation tasks under two views is the substantial heterogeneity observed in the two types of information, evident in both their data structures and the articulation of user interests. To surmount this challenge, this paper capitalizes on the effectiveness of structured data in amalgamating diverse information sources \cite{ji2015knowledge, zhu2019multi, zhou2022tale, qiu2022contrastive}. Rather than directly learning the synergistic relationship between the two raw information types, our approach prioritizes the initial construction of highly structured interest representations for each view, upon which we subsequently build the understanding of the synergistic relationship between the two information views.

Specifically, in this paper, we propose a novel framework, termed \textbf{Bi}directional \textbf{V}iew-based Multimodal Sequential \textbf{Rec}ommendation (\textbf{BivRec}), consisting of three key components:
(i) Multi-scale Embedding Block: this block is designed to extract user interests comprehensively from each view by employing multi-scale patching operations on the user interaction sequences, facilitating a nuanced understanding of user preferences across different time scales.
(ii) Intra-view Decoupling Block: this block then incorporates our proposed Gaussian Attention and Cluster Attention mechanisms, thereby generating highly structured interest representations. The Gaussian Attention mechanism captures feature interactions with temporal dependencies between user interactions, while the Cluster Attention mechanism conducts trainable clustering operations within each view.
(iii) Cross-view Interest Learning Block: finally, we leverage both overall interest semantic information and the interest allocation matrix to learn the synergistic relationship between user interests in the two views. This block fosters mutual reinforcement and refinement of recommendations from both ID and multimodal perspectives.
Our contributions can be succinctly summarized as follows:
\begin{itemize}[leftmargin=*]
    \item To the best of our knowledge, this is the first research that proposes a joint training framework for recommendation tasks encompassing both ID and multimodal views, addressing the limitations inherent in existing MSR frameworks; 
    \item To tackle the challenges arising from information heterogeneity in joint learning, we present a novel recommendation method named BivRec, comprising three modules for interest modeling, which collectively provide a robust solution to the complexities of multimodal recommendation;
    \item Empirical results on five public benchmark datasets underscore the superiority of BivRec over baseline models, demonstrating its exceptional performance across various scenarios.
    \vspace{-1mm}
\end{itemize} 
\vspace{-1mm}

\section{Framework}

\begin{figure*}[htp]
    \centering
    \includegraphics[width=0.95\linewidth]{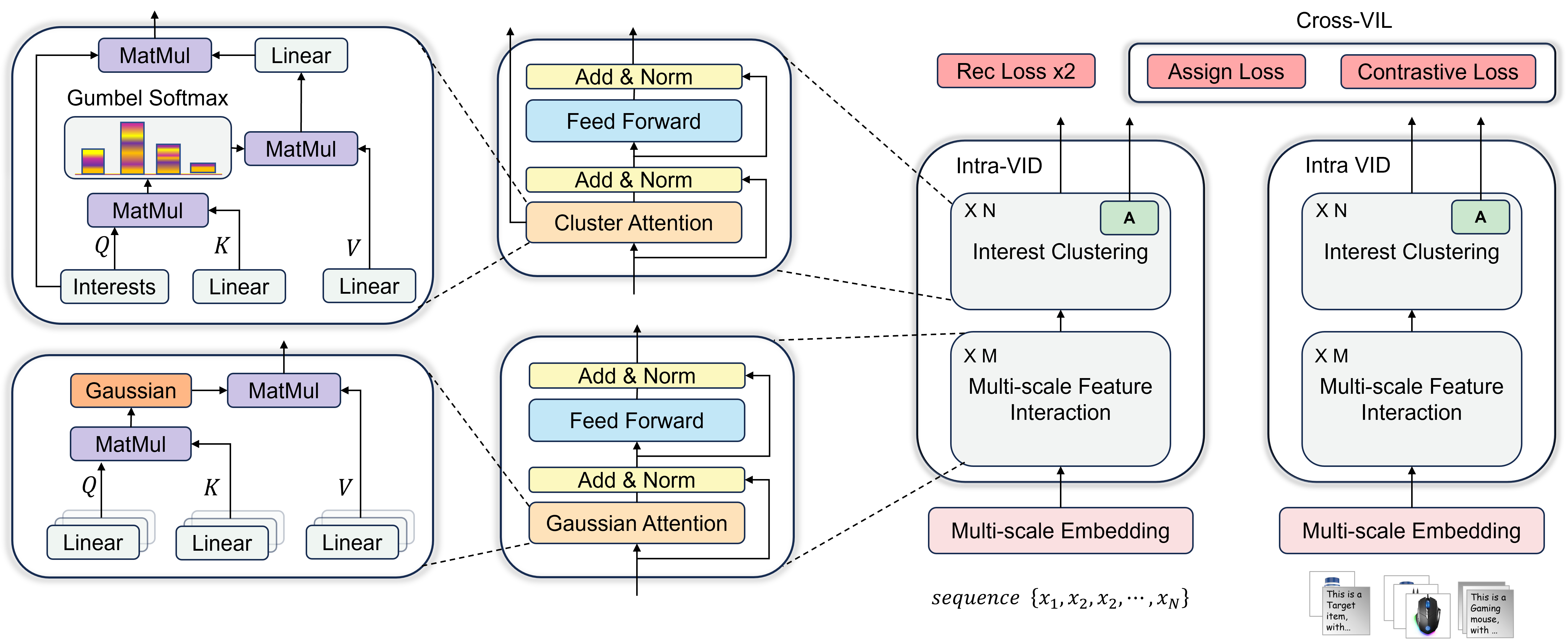}
    \vspace{-3mm}
    \caption{The overview framework of BivRec.}
    \label{fig:model}
    \vspace{-2mm}
\end{figure*}

In this section, a detailed description of the proposed BivRec framework is provided. The problem formulation of the sequential recommendation task under two views in a multi-embedding paradigm is presented, followed by a comprehensive introduction of the proposed Multi-scale Embedding Block, Intra-View Interest Decomposition Block (Intra-VID), and Cross-View Interest Learning Block (Cross-VIL). Subsequently, the model optimization process is elaborated and the corresponding pseudo-code is provided.


\subsection{Problem Formulation}

The recommendation task under ID and multimodal views is performed within a multi-scale and multi-embedding paradigm.

\noindent \textbf{ID View:} For user $\boldsymbol{u}$, her/his interaction sequence could be represented as $\boldsymbol{S}_{u}=\left\{\boldsymbol{x}_{1}, \cdots, \boldsymbol{x}_{i}, \cdots, \boldsymbol{x}_{N}\right\} $ including $N$ items. The multi-scale embedding is then applied to get $\boldsymbol{E}_u^{id}\in \mathbb{R}^{N^{\prime} \times D^{id}}$, where $N^{\prime}$ is the new sequence length after multi-scale expanding and $\boldsymbol{D}^{id}$ is the dimension of hidden state. 

\noindent \textbf{Multimodal view:} A similar multimodal embedding $\boldsymbol{E}_{u}^{m} \in \mathbb{R}^{N^{\prime} \times D^{m}}$ could be constructed for $\boldsymbol{S}_{u}$ based on its corresponding multimodal features, where $\boldsymbol{D}^m$ is dimension of multimodal feature. For simplicity, we use $m$ to represent multimodal features. In practice, either one or both visual and textual features could be utilized. If both are used, their features will be concatenated to create a joint representation.

Based on the two above views, we aim to construct two dense interest matrices $\boldsymbol{V}_u^{id} \in \mathbb{R}^{K \times D^{id}}$ and $\boldsymbol{V}_{u}^{m} \in \mathbb{R}^{K \times D^m}$ in both views, where ${K}$ is a pre-defined number of interests, $\boldsymbol{v}_{k}^{id}$ and $\boldsymbol{v}_{k}^m$ is for each interest.
Then the problem of bidirectional view-based multimodal sequential recommendation can be formally defined as: \textit{Given the user interacted item sequence including $N$ time steps, our goal is to predict the $N + 1^{th}$ item based on $\boldsymbol{V}_u^{id}$ and $\boldsymbol{V}_u^m$ separately.}

Note that, (i) in this paper, we omit the subscript $u$ for simplicity, and (ii) as the recommendation tasks in both views utilize the same model structure, we only provide the formulas for the recommendation task in the ID view as an example.

\subsection{Framework Overiew}
As shown in Figure \ref{fig:model}, BivRec is a dual-tower symmetric structure with the same architecture. In each view, the user interaction sequences are first transformed into multi-scale embedding representations. Subsequently, these representations undergo the Intra-View Interest Decomposition Block (Intra-VID) with a Multi-scale Feature Interaction Layer and Interest Clustering Layer, exporting structured interest representations and the interest assignment matrices. Additionally, the Assign loss and Contrastive loss in the Cross-View Interest Learning block (Cross-VIL) with two recommendation losses are conducted to optimize the model. 

\vspace{-2mm}

\subsection{\textbf{Multi-scale Interest Embedding}}
In current sequence recommendation frameworks, a common approach is to utilize shared embedding matrices to obtain distinct vector representations for each item. These representations are then used to derive user representations based on their historical interaction sequences. However, in real-world recommendation scenarios, a user's purchasing behavior for the next item can be influenced by a combination of multiple items. For example, consider a user who purchases office supplies, such as a portable hard drive, a printer, and a mouse. To enhance the office setup, he decides to purchase an ergonomic chair as the next item. In this case, understanding user interests requires considering these three aspects as a whole, as a single item alone is not sufficient.

To address the above challenge, a multi-scale user interest embedding is incorporated in BivRec.
Specifically, we expand the user interaction sequence by segmenting it into different scales and then utilize additive concatenation to aggregate item embeddings for each scale $s$. In addition, two learnable positional embeddings are introduced to annotate scale information and interaction sequence position information.

For user embedding in ID view, initially, an original item embedding $\boldsymbol{F}\in \mathbb{R}^{N\times D^{id}}$ could be acquired through a trainable embedding table and a sequence position embedding $\boldsymbol{E}_{pos}\in \mathbf{R}^{N \times D^{id}}$ could be added to get user interest embedding in scale 1, noted as $\boldsymbol{E}_1^{id}$:
\begin{equation}
\boldsymbol{E}_1^{id}=\boldsymbol{F} + \boldsymbol{E}_{pos}.
\end{equation}
Based on $\boldsymbol{E}_1^{id}$, the user interest embedding in scale $s$ could be constructed as:
\begin{equation}
\boldsymbol{E}_s^{id} = merge\left(\text{Unfold}\left({E}_1^{id} , kernel=s , stride=s\right)\right) + \boldsymbol{P}_{s},
\end{equation}
where $\boldsymbol{P}_{s}\in \mathbb{R}^{N/s\times D^{id}}$ is scale position embedding, Unfold function conducts non-overlapping patching operation on $\boldsymbol{E}_1^{id}$, $merge$ is a sum operation in the scale dimension. In this scenario, each scale's embedding is accompanied by independent scale position embedding and shared sequence position embedding.

Finally, we concat the embeddings in totally $S$ scales to obtain multi-scale user interest embedding $\boldsymbol{E}^{id}\in\mathbf{R}^{N^{\prime}\times D^{id}}$:
\begin{equation}
\boldsymbol{E}^{id} = [\boldsymbol{E}_1^{id},\cdots ,\boldsymbol{E}_s^{id}, \cdots, \boldsymbol{E}_S^{id}]^\top.
\end{equation}

\subsection{Intra-View Interest Decomposition}
Most existing models primarily utilize self-attention mechanisms to adaptively model user interests. However, this approach often overlooks the temporal dependencies present in the interaction sequence during feature interactions and conducts non-structured user interests \cite{zhang2022re4}. 

To address these limitations, we adhere to a stackable framework to design an Intra-VID Block consisting of $M$ Multi-scale Feature Interaction Layer and $N$ Interest Clustering Layer.
 The number of layers will depend on the length of the dataset interaction records. In this block, the Multi-scale Feature Interaction Layer focuses on pre-information passing, and the Interest Clustering Layer is used to conduct structured interest representations.

\subsubsection{\textbf{Multi-scale Feature Interaction}}
The computation of each value in the attention matrix under the standard attention mechanism \cite{vaswani2017attention} in $l^{th}$ layer is revisited. For simplicity, we omit the multi-head mechanism in our description:
\begin{equation}
\bar{\boldsymbol{A}}^l_{i j}=\boldsymbol{W}_{q}^l \boldsymbol{x}_{i}^l \cdot \boldsymbol{W}_{k}^l \boldsymbol{x}_{j}^l,
\label{attn}
\end{equation}
where $\boldsymbol{W_q}$ and $\boldsymbol{W_k}$ are the weights of the learned linear projections, $\boldsymbol{x}^l$ is the token in the sequence. However, in this case, relying solely on vector representations for similarity computation overlooks a crucial aspect: the relationships between items in a user interaction sequence are not entirely equal. Generally, items that are closer in interaction sequence exhibit higher correlations. We aim to enable the model to adapt to this position-dependent relationship when assigning weighted distributions to interest information. Specifically, a Gaussian function is introduced and it could exponentially decrease the relationship between tokens as their positions become further apart:
\begin{equation}
\begin{aligned}
    \boldsymbol{G}_{ij}^l = \exp \left(-\frac{\boldsymbol{D}_{ij}^2}{{\sigma^l}^2}\right) / \sum_{j=1}^{N^{\prime}} \exp \left(-\frac{\boldsymbol{D}_{ij}^2}{{\sigma^l}^2}\right),
\end{aligned}
\end{equation}
where $\boldsymbol{D}\in\mathbf{R}^{N^{\prime}\times N^{\prime}}$ is the distance matrices for each token. Due to the inclusion of multi-scale information in our embedding sequence, for token representations with a scale greater than 1, we replace them with the average of the positional information of the original item embeddings they contain. $\sigma$ is used to control the constrained area, in our model, it is a trainable parameter. Besides, normalization operations are also applied in the last dimension.

So in Gaussian attention, Eq. \ref{attn} can be rewritten as:
\begin{equation}
\boldsymbol{A}_{i j}^{l}=\boldsymbol{G}_{ij}^l \cdot \boldsymbol{W}_{q}^l \boldsymbol{x}_{i}^l \cdot \boldsymbol{W}_{k}^l \boldsymbol{x}_{j}^l.
\end{equation}

\subsubsection{\textbf{Interest Clustering}}
In each Interest Clustering Layer, $K^l$ trainable interest tokens $\{\boldsymbol{c}_i^l\}_{i=1}^{K^l}$ are initialized by truncated distribution to represent different aspects of user interests. In practice, it is common to have only one Interest Clustering layer, with a single attention head. Only for long interaction sequences, a multi-stage interest clustering is performed. We exemplify the clustering attention mechanism with single-head and single-layer, using $\boldsymbol{c}_i$ as the query, multi-scale item embedding token $\boldsymbol{x}_j$ as the key to computing each attention score for interest assignment:
\begin{equation}
\boldsymbol{A}_{i j}^c=\boldsymbol{W}_{q}^c \boldsymbol{c}_{i} \cdot \boldsymbol{W}_{k}^c \boldsymbol{x}_{j},
\end{equation}
where $\boldsymbol{W}_q^c$ and $\boldsymbol{W}_k^c$ are the weights of the learned linear projections for the interest and item tokens in the Interest Clustering Layer. 

To construct highly structured interest representations, a Filter operation is introduced with a hyperparameter $F$ along the dimension
of interest tokens:

\begin{equation}
filter(\boldsymbol{A}_{i,j}) = 
\begin{cases}
1, &\text{if} ~~~ \boldsymbol{A}_{i,j} = \max\limits_{k \in topF(i)} \boldsymbol{A}_{i,k} \\ 
0, &\text{otherwise}
\end{cases},
\label{Allocation}
\end{equation}
where the $filter$ is the Filter operation. $topF$ means each item assigns interest information to the $F$ proximate interest tokens to ensure that the interests are strictly from different aspects, thereby achieving structured representation.
However, the $top$ operation is non-differentiable and leads to a non-exploratory network. So, the Gumbel softmax \cite{jang2016categorical, xu2022groupvit} is applied instead of softmax operation in computing the attention matrix:

\begin{equation}
\boldsymbol{A}_{i j}^{c'}=\frac{\exp \left(\left(\boldsymbol{A}_{i,j}^c+ \gamma_{i}\right)\right)}{\sum_{k=1}^K \exp \left(\left(\boldsymbol{A}_{i,j}^c+ \gamma_{k}\right) \right)},
\end{equation}
where $\gamma$ is i.i.d random samples drawn from the $Gumbel$ (0, 1)\footnote{The $Gumbel$ (0, 1) distribution can be sampled using inverse transform sampling by drawing $u$ from $Uniform$ (0, 1) and computing $\gamma=-log(-log(u))$.} distribution, which is controlled by a temperature coefficient $\tau$. For low temperatures, the expected value of a Gumbel softmax random variable approaches the expected value of a categorical random variable with the same logits. On the other hand, the expected value converges to a uniform distribution over the categories as the temperature increases. 

After gaining the attention matrix, a straight-through trick  \cite{van2017neural} is utilized to form the final attention matrix:
\begin{equation}
\boldsymbol{\hat{A}}=filter\left(\boldsymbol{A}_{i j}^{c'}\right) + \boldsymbol{A}_{i j}^{c'} - sg\left(\boldsymbol{A}_{i j}^{c'}\right),
\end{equation}
where $sg$ is a stop gradient operation. In this way, when the network propagates forward, it implements the max $F$ operation to construct explicit structured interests. At the same time, when the network goes backward to update the parameter, it has the same gradient as $\boldsymbol{A}$ to maintain exploratory capabilities. The computational complexity of cluster attention is approximately linear due to a fixed number of interests $K$ (usually 4).

Now, the structured multi-interest representations for recommendation tasks could be constructed. It is achieved by assigning the multi-scale item
embedding token $\boldsymbol{x}_{j}$ to the interest token $\boldsymbol{c}_{k}$ through attention score $\boldsymbol{\hat{A}}_{ij}$ and adding a residual structure:
\begin{equation}
    \boldsymbol{V}^{id} = [\boldsymbol{v}_{1}^{id}, \boldsymbol{v}_{2}^{id}, \cdots, \boldsymbol{v}_{k}^{id} ]^{\top},
\end{equation}
\vspace{-3mm}
\begin{equation}
    \boldsymbol{v}_{k}^{id} = \boldsymbol{c}_{k} + \boldsymbol{W}_{a}\hat{\boldsymbol{A}} \boldsymbol{W}_{v} \boldsymbol{x}_{j},
\end{equation}
where $\boldsymbol{V}^{id}$ is the final structured interests representations of user $\boldsymbol{u}$ in ID view, $\boldsymbol{v}_{k}^{id}$ is the $k^{th}$ interest representation of user interests. $\boldsymbol{W}_{a}$ and $\boldsymbol{W}_{v}$ are learned weights to control the interest information to allocate to interest tokens. So far, we have constructed structured representations of interest, the principles of which will be further elaborated in the experimental section.

\vspace{-1mm}

\subsection{Cross-View Interest Learning}
Due to the highly heterogeneous nature of interest representations $\boldsymbol{V}^{id}$ and $\boldsymbol{V}^m$, which contain multiple interests in both views, it is difficult to establish a one-to-one correspondence between user interests under two views. In this section, a detailed explanation of how BivRec learns the synergistic relationship between the structured interest representations by both coarse-grained and fine-grained is provided.
\subsubsection{\textbf{Coarse-grained Learning}}
For the recommendation task in multimodal view, a similarly Multi-scale Interest Embedding Block $f_{MIE}^{m}$ and Intra-VID Block $f_{VID}^{m}$ are used to construct the same number of interest representations $\boldsymbol{V}^m$:
\begin{equation}
    \boldsymbol{V}^m = [\boldsymbol{v}_{1}^m, \boldsymbol{v}_{2}^m, \cdots, \boldsymbol{v}_{k}^m ] = f_{VID}^{m}\left(f_{MIE}^{m}\left(\boldsymbol{S}^m\right)\right).
    \label{get interest}
\end{equation}
A trainable average pooling layer is initially utilized to get a unit semantic presentation of user interests under each view to learn the overall semantic similarity: 
\vspace{-1mm}
\begin{equation}
\begin{aligned}
\boldsymbol{h}^{id} &= \boldsymbol{f}_{{AVG}^{ID}}\left(\boldsymbol{V}^{id}\right), \\
\vspace{-2mm}
\boldsymbol{h}^m &= \boldsymbol{f}_{{AVG}^{m}}\left(\boldsymbol{V}^m\right),
\end{aligned}
\end{equation}
where $\boldsymbol{h}^m$ and $\boldsymbol{h}^{id}$ are two vectors $\in$ $\mathbb{R}^{D}$ in D dimension, representing all the interest information under multimodal and ID view. $\boldsymbol{f}_{AVG}$ is the Average pooling layer. Then, two MLP layers are applied to embed $\boldsymbol{h}^m$ and $\boldsymbol{h}^{id}$ into a common embedding space as $\hat{\boldsymbol{h}}^{id}$ and $\hat{\boldsymbol{h}}^m$ through nonlinear transformations:
\begin{equation}
\label{unit}
\begin{aligned} 
\hat{\boldsymbol{h}}^{id} &= \boldsymbol{W}_1^{id} GELU \left(\bold{W}_0^{id} \boldsymbol{h} + \boldsymbol{b}_0^{id}\right)+\boldsymbol{b}_1^{id}, \\
\hat{\boldsymbol{h}}^m &= \boldsymbol{W}_1^m GELU \left(\bold{W}_0^m \boldsymbol{h}^m + \boldsymbol{b}_0^m\right)+\boldsymbol{b}_1^m,
\end{aligned}
\end{equation}
where $GELU$ is the activate function, all the $\boldsymbol{W}$ and $\boldsymbol{b}$ are trainable matrixes and bias.

We use InfoNCE \cite{oord2018representation} loss function to learn the overall semantic similarity in a batch of interest pair $\{\hat{\boldsymbol{h}}^{id}, \hat{\boldsymbol{h}}^m \}_{i=1}^B$. Our goal is to ensure that the interest representations of the same user under the two views represent the same semantic information about their interests. To simplify the model, data augmentation like \cite{xie2022contrastive, xu2022groupvit} is not performed. Specifically, for each pair of interest representations, the two interest representations belonging to the same user are treated as positive pairs, and those belonging to different users are treated as negative pairs. We pull the distance between positive interest representations closer and push away the distance between negative interest representations in the common embedding space. We set dual direction view based contrastive loss to learn interest information synergistic relationship from both multimodal to ID view and ID to multimodal view \cite{xu2022groupvit}:
\begin{equation}
\mathcal{L}_{MM \leftrightarrow ID}=\mathcal{L}_{MM \rightarrow ID}+\mathcal{L}_{ID \rightarrow MM},
\end{equation}
For example, the loss function from multimodal view to ID view is:
\begin{equation}
\mathcal{L}_{MM \rightarrow ID}=-\frac{1}{B} \sum_{i=1}^B \log \frac{\exp \left(\hat{\boldsymbol{h}_{i}}^m \cdot \hat{\boldsymbol{h}_{i}^{id}} / \beta\right)}{\sum_{j=1}^B \exp \left(\hat{\boldsymbol{h}_{i}}^m \cdot \hat{\boldsymbol{h}_{j}}^{id} / \beta\right)},
\end{equation}
where $\beta$ is a learnable temperature parameter to scale the logits. So far, we preliminarily express the interest information in two views through the interest-structured blocks. Then, we conduct self-supervised learning to learn the synergistic relationship of interest information between two views.

\subsubsection{\textbf{Fine-grained Learning}}
Apart from learning the overall interest semantic similarity, we aim to enable the model to learn fine-grained similarity in interest allocation. This can be achieved by comparing the interest allocation matrices derived from two distinct perspectives, as demonstrated in Eq. \ref{Allocation}.
In the interest information allocation matrices for each view, rows indicate multi-scale item information allocated to the same interest vector, while columns represent the allocation of multi-scale item information to specific interests. Due to the high heterogeneity between the two types of information, the interest representations do not have a one-to-one correspondence. Therefore, we focus on evaluating the similarity between rows of the allocation matrices. The loss function can be formulated accordingly:
\begin{equation}
\mathcal{L}_{A} = - \sum_{k=1}^K cos (\boldsymbol{A}_A^{m}[k,:]\cdot \boldsymbol{A}_A^{id}[k,:]),
\end{equation}
where $cos$ is the cosine similarity, $\boldsymbol{A}_A^{m}$ and $\boldsymbol{A}_A^{id}$ are the allocation matrices under two view.

\subsection{Optimization}
Both recommendation tasks in this paper follow the loss of the multi-embedding sequential recommendation framework \cite{cen2020controllable} during the optimization process. Taking the recommendation task in the ID view as an example, the final interest representations $\hat{\boldsymbol{V}_u}^{id}$ (represented by $\hat{\boldsymbol{V}}$) could be obtained for recommending tasks. Then, an $\arg \max$ operator is used to choose the closest user interest vector $\hat{\boldsymbol{v}}$ for the ${N+1}^{th}$ item embedding $\boldsymbol{x}{N+1}$ at the timestep:
\begin{equation}
\hat{\boldsymbol{v}}=\hat{\boldsymbol{V}}\left[:~, \operatorname{argmax}\left(\hat{\boldsymbol{V}}^{\top} \cdot \boldsymbol{x}_{N+1}\right)\right].
\end{equation}
 Given a training sample $\boldsymbol{v}$ and
$\boldsymbol{x}_{N+1}$, the likelihood of the user $\boldsymbol{u}$ interacting with the item could be computed as:
\begin{equation}
P_{\theta} (\boldsymbol{t} | \boldsymbol{i}) = \frac{\exp \left(\boldsymbol{v}^{\top} \cdot \boldsymbol{x}_{N+1}\right)}  {\sum_{k \in K} \exp \left(\boldsymbol{v}^{\top} \cdot \boldsymbol{x}_k\right)}.
\end{equation}
The objective function of our model is to minimize the following
negative log-likelihood: 
\begin{equation}
\mathcal{L}_{Rec}=\sum_{\boldsymbol{u} \in \mathcal{U}} \sum_{\boldsymbol{t} \in \mathcal{I}}-\log P\left(\boldsymbol{i}| \theta \right).
\label{recloss}
\end{equation}
The sum operator is computationally expensive. Thus, a sampled softmax technique \cite{covington2016deep} is applied to train our model. The final training loss function is a multi-task loss:
\begin{equation}
\mathcal{L}=\mathcal{L}_{Rec}^{ID} + sg (\frac{\mathcal{L}_{Rec}^{ID} }{\mathcal{L}_{Rec}^{MM}}) \mathcal{L}_{Rec}^{MM} +\lambda_1 \mathcal{L}_{MM\leftrightarrow ID} + \lambda_2 \mathcal{L}_{A},
\label{loss}
\end{equation}
where $sg$ is the stop gradient operation used to eliminate numerical effects and ensure equal importance of recommendation tasks in both views. $\lambda$ is the hyper-parameter.


\begin{table*}[t]
\centering
\caption{The overall performance of different models on five datasets. Models in bold indicate our proposed model. Scores in bold indicate the best model for each metric, and scores underlined indicate the second-best model, except for our proposed model.}
\vspace{-3mm}
\setlength{\tabcolsep}{4pt}
\resizebox{2\columnwidth}{!}{
\begin{tabular}{c|ccc|ccc|ccc|ccc|ccc}
\toprule[1.5pt]
\multicolumn{1}{c}{\textbf{\Large Model}}&\multicolumn{15}{c}
{\textbf{\Large Multimodal Recommendation Dataset}}\\
\cmidrule(l){1-1}\cmidrule(l){2-7}\cmidrule(l){8-16}
\textbf{Metrics@20}&\multicolumn{3}{c}{Ml-25m}&\multicolumn{3}{c}{Ml-1m}&\multicolumn{3}{c}{Elec}&\multicolumn{3}{c}{Cloth}&\multicolumn{3}{c}{Baby}\\
\cmidrule(l){2-4}\cmidrule(l){5-7}\cmidrule(l){8-10}\cmidrule(l){11-13}\cmidrule(l){14-16}
\textbf{Units \% } &Recall & \ \ \ NDCG &HR &Recall & \ \ \ NDCG &HR &Recall & \ \ \ NDCG &HR &Recall & \ \ \ NDCG &HR &Recall & \ \ \ NDCG &HR\\
\midrule
\midrule
\multirow{1}{*}{ComiRec-SA} 
&5.41 &17.22 &48.15 &7.46 & 24.42& 58.78& 4.23& 2.14& 6.37& 1.47& 0.81& 2.45& 4.24& 3.57 & 7.69\\
\midrule
 \multirow{1}{*}{SASRec*} 
&5.24 & 16.40& 48.55& 7.32& 24.04 & 58.63 & 4.28 & 2.22& 6.40& 1.44& 0.77& 2.34 & 4.19 & 3.49 & 8.03\\
\midrule
\multirow{1}{*}{ComiRec-SA+} 
&6.16 & 19.23 & 50.22 & 7.59& 24.98& 59.34& 5.18& 2.96& 8.01 &2.13 & 1.19 & 3.04 & 4.69 & 3.80 & 8.97 \\
\midrule
\multirow{1}{*}{SASRec*+} 
&6.33 &18.94 & 51.56 & 8.01 & 25.81 & 60.04 & 5.22 & 3.29 & 8.78 & 1.80 & 1.12 & 2.98 & 4.75 & 3.92 & 9.22 \\
\midrule
\multirow{1}{*}{FDSA*+} 
& 6.98 & 20.03 & 54.24 & 8.23 & 26.63 & 60.74 & 5.04 & 3.63 & 9.93 & 2.08 & 1.53 & 3.14 & 5.19 & 4.22 & 9.97\\
\midrule
\multirow{1}{*}{MMSRec*} 
& 6.52 & 19.65 & 52.73 & 8.15 & 25.92 & 59.77 & 5.01 & 3.24 & 8.38 & 1.92 & 1.41 & 2.99 & 4.61 & 3.95 & 8.86\\
\midrule
\multirow{1}{*}{MMMLP*} 
& 7.38 & 22.09 & 56.17& 8.64 & 26.93& 62.09 & 5.12 & 4.21& 10.36 & 2.41 & 1.98 & 3.81 & 5.80 & 4.41 & 10.32 \\
\midrule
\multirow{1}{*}{NOVA*+} 
&{8.37} &{24.37}& {60.20}& {9.19}& {28.84}& {63.41}& {5.30}& {5.17}&{11.49} &{2.97} & {2.34}& {4.64}& {6.11} &{4.92} & {12.09}\\
\midrule
\multirow{1}{*}{DIF-SR*+} 
& 8.91 & 25.29 & 59.42& 10.01 & 29.17& 63.09 & \underline{5.51} & 5.44& 11.96 & \underline{3.15} & \underline{2.39} & 4.74 & \underline{6.27} & \underline{5.01} & \underline{12.22} \\
\midrule
\multirow{1}{*}{\textbf{BivRec-ID}}
&\underline{9.41} & \underline{26.97} & \underline{63.20} & \underline{10.04} & \underline{29.26} & \underline{64.88} & 5.34& \underline{5.55}& \underline{12.27} &2.95 & 2.11& \underline{4.81} &6.18 & 4.82 &12.08\\
\midrule
\multirow{1}{*}{\textbf{BivRec-MM}}
&\textbf{11.96*} &\textbf{28.31*} &\textbf{65.54*} &\textbf{12.50*} &\textbf{32.36*} &\textbf{66.86*} & \textbf{5.86*} & \textbf{5.93*} & \textbf{13.98*} &\textbf{3.86*} &\textbf{2.61*} &\textbf{5.77*} &\textbf{6.50*} &\textbf{5.97*} &\textbf{13.05*}\\
\midrule
\multirow{1}{*}{Improve} 
 & 34.23\%  & 11.94\% & 10.30\% & 24.88\% &10.94\% & 5.44\%& 6.35\% & 9.01\% & 16.89\% & 22.54\% & 9.21\% &21.73\% &3.67\% & 17.10\% & 6.79\% \\
\bottomrule[1.5pt]
\toprule[1.5pt]
\multirow{1}{*}{\textbf{PureID}}
 & 9.20  & 24.70 & 60.10 & 9.04 & 27.12 & 63.10 & 5.43& 3.89 & 8.96& 2.29 & 1.83 & 4.19 & 6.05 & 4.44 &11.25\\
\midrule
\multirow{1}{*}{\textbf{BivRec-ID}}
&9.41 & 26.97 &63.20 & 10.04 & 29.26 & 64.88 & 5.14& 5.55& 12.27 &2.95 & 2.11& 4.81 &6.18 & 4.82 &12.08\\
\midrule
\multirow{1}{*}{\textbf{PureMM}}
 & 11.62  & 28.01 & 64.64 & 12.34 & 30.92 & 65.19 & 5.76& 5.91 & 13.79& \textbf{4.01} & \textbf{2.84} & \textbf{6.28} & 6.47 & 5.47 &12.16\\
\midrule
\multirow{1}{*}{\textbf{BivRec-MM}}
&\textbf{11.96} &\textbf{28.31} &\textbf{65.54} &\textbf{12.50} &\textbf{32.36} &\textbf{66.86} & \textbf{5.86} & \textbf{5.93} & \textbf{13.98} &{3.86} &{2.61} &{5.77} &\textbf{6.50} &\textbf{5.97} &\textbf{13.05}\\
\bottomrule[1.5pt]
\end{tabular}}
\\``\textbf{{\Large *}}'' in scores indicates the statistically significant improvements (i.e., two-sided t-test with $p<0.05$) over the best baseline.
\label{tab:overview}
\vspace{-3mm}
\end{table*}

\section{Experiment}
In this section, we conduct experiments to demonstrate the validity of our proposed framework. For details of experimental settings, dataset descriptions, and baseline model specifications, please refer to Appendix \ref{setting}. Furthermore, we also investigate the performance of BivRec when applied to the recall stage on two large scale datasets in Appendix \ref{recall stage} and a case study in Appendix \ref{case}. Specifically, our experiments aim to answer the following questions:
\vspace{-1mm}
\begin{itemize}[leftmargin=*]
    \item \textit{\textbf{RQ 1:}} How does BivRec compare to existing multimodal sequential recommendation models?
    \item \textit{\textbf{RQ 2:}} Whether BivRec can improve the recommendation performance bidirectionally?
    \item \textit{\textbf{RQ 3:}} Whether each block in addressing information heterogeneity under two views is effective and how does it achieve this?
    \item \textit{\textbf{RQ 4:}} Does BivRec inherit the advantages of pure multimodal recommendation models?
\end{itemize}
\vspace{-1mm}

\begin{table}[t]
        \centering
	\caption{Module Ablation in ML-1m. \textit{w/o} means without.}
        \vspace{-3mm}
        \setlength{\tabcolsep}{4pt}
	\label{table:each block}
        \resizebox{1\columnwidth}{!}{
	\begin{tabular}{lccc}
		\toprule[1.5pt]
		\textbf{{Model Architecture}} & \textbf{{ID}} & \textbf{{MM}}  \\
        \midrule
        \midrule
        \ {Single branch in BivRec}  & \textbf{9.04} & \textbf{12.34}  \\
        \ {\textit{w/o} Multi-scale Interest Embedding}  & 8.87 (-3.59\%) & 11.62 (-5.83\%)  \\
		\ {\textit{w/o} Gaussian Attention}  & 8.65 (-4.31\%) & 11.58 (-6.16\%)   \\
        \ {\textit{w/o} Cluster Attention} & 8.12 (-10.18\%) & 10.23 (-17.10\%) \\
        \midrule
        \ {BivRec} &\textbf{10.04} &\textbf{12.50}  \\
        \ {\textit{w/o} Multi-scale Interest Embedding}  & 9.72 (-3.19\%) & 12.11 (-3.12\%)  \\
		\ {\textit{w/o} Gaussian Attention}  & 9.60 (-3.59\%) & 11.93 (-3.59\%)   \\
        \ {\textit{w/o} Cluster Attention} & 7.98 (-20.52\%) & 9.42 (-24.64\%) \\
        \ {\textit{w/o} $\mathcal{L}_{A}$ } &9.75 (-2.89\%) &11.68 (-6.56\%)  \\
        \bottomrule[1.5pt]
	\end{tabular} 
}
\vspace{-1em}
\end{table}

\vspace{-2mm}
\subsection{Overall Performance Comparison (RQ1, RQ2)}

To answer \textbf{RQ1}, we conduct an overall performance comparison. BivRec-ID and BivRec-MM jointly train the recommendation task in two views, utilizing the ID and multimodal information for the recommendation. The top part of Table \ref{tab:overview} shows that 
(i) All multimodal versions of the model outperform the original model, showing the importance of incorporating multimodal information in modeling user interests.
(ii) Our proposed BivRec achieves the best performance in all five datasets, demonstrating the performance of joint training. BivRec-MM performs better, indicating that interest information is richer in multimodal information.
(iii) MMSRec directly utilizes multimodal information for recommendations, but it may not exhibit significant performance improvements compared to SASRec. In fact, it may experience performance degradation on datasets like Elec and Baby. This highlights the effectiveness of BivRec in constructing structured interest representations.
(iv) NOVA and DIF-SR have successfully addressed the issue of information structure intrusion, leading to significant performance improvements compared to other multimodal sequence recommendation baseline models. This highlights the importance of BivRec's adoption of dual perspectives in the recommendation process.

To answer \textbf{RQ2}, we also test the performance of recommendation models individually trained under the ID view (PureID) and MM view (PureMM). As shown in the lower part of Table \ref{tab:overview}, we can observe that BivRec-ID outperforms PureID in all datasets. BivRec-MM outperforms PureMM except for the cloth dataset. This demonstrates that the joint training approach can make the learned synergistic relationship improve recommendation performances in both views. It is possible that the multimodal features of clothing items play a dominant role in influencing a user's purchase decision, which may explain the lack of improvement in the clothing dataset. Besides, by exploring the synergistic relationship of interests within the two views, it becomes evident that the performance improvement of recommendation tasks is more pronounced in the ID view compared to the multimodal view.

\begin{figure}[t]
	\subfloat[different modality]
    {\includegraphics[width=.44\columnwidth]{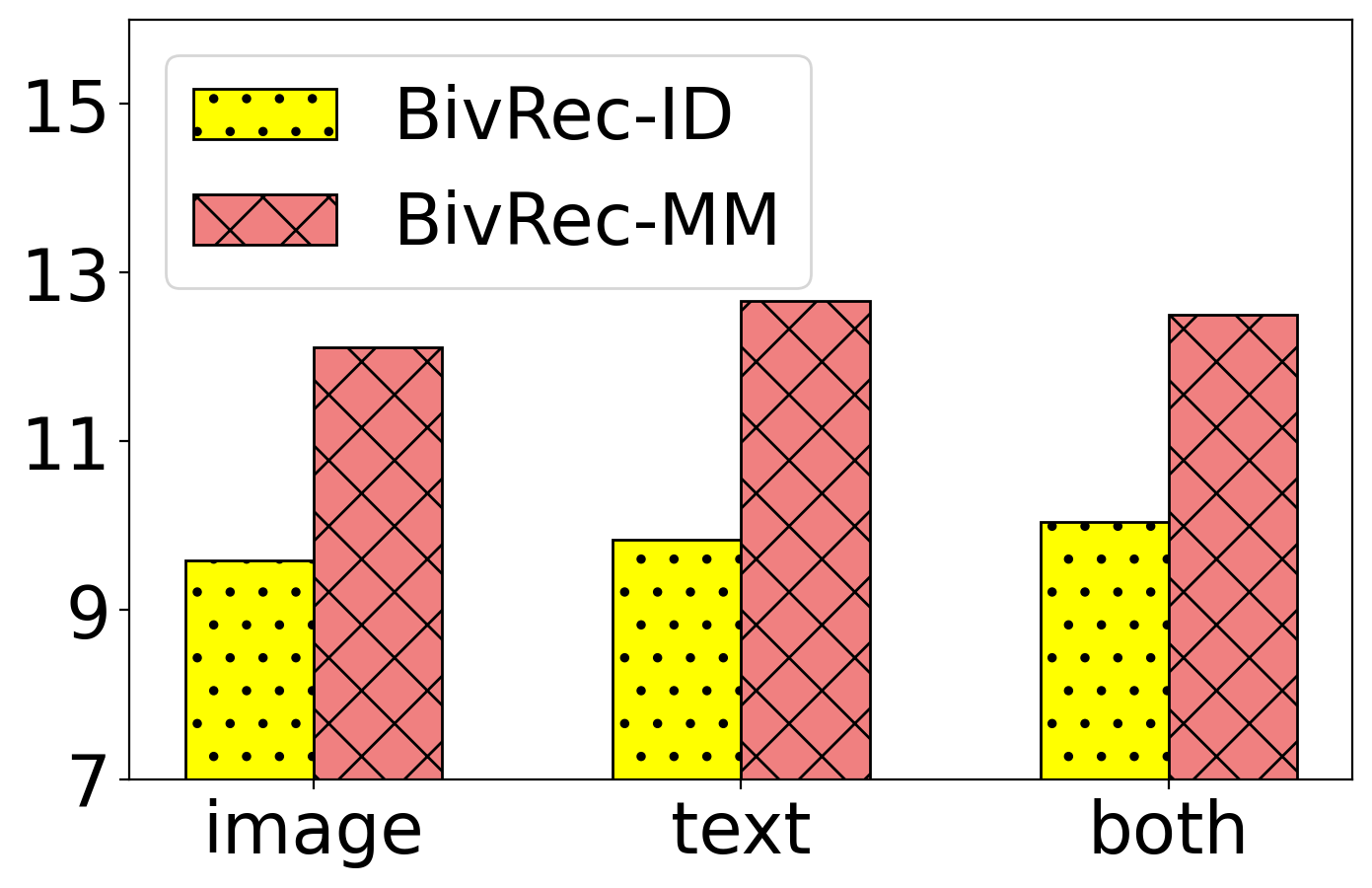}}\hspace{4pt}
	\subfloat[different info allocation]{\includegraphics[width=.44\columnwidth]{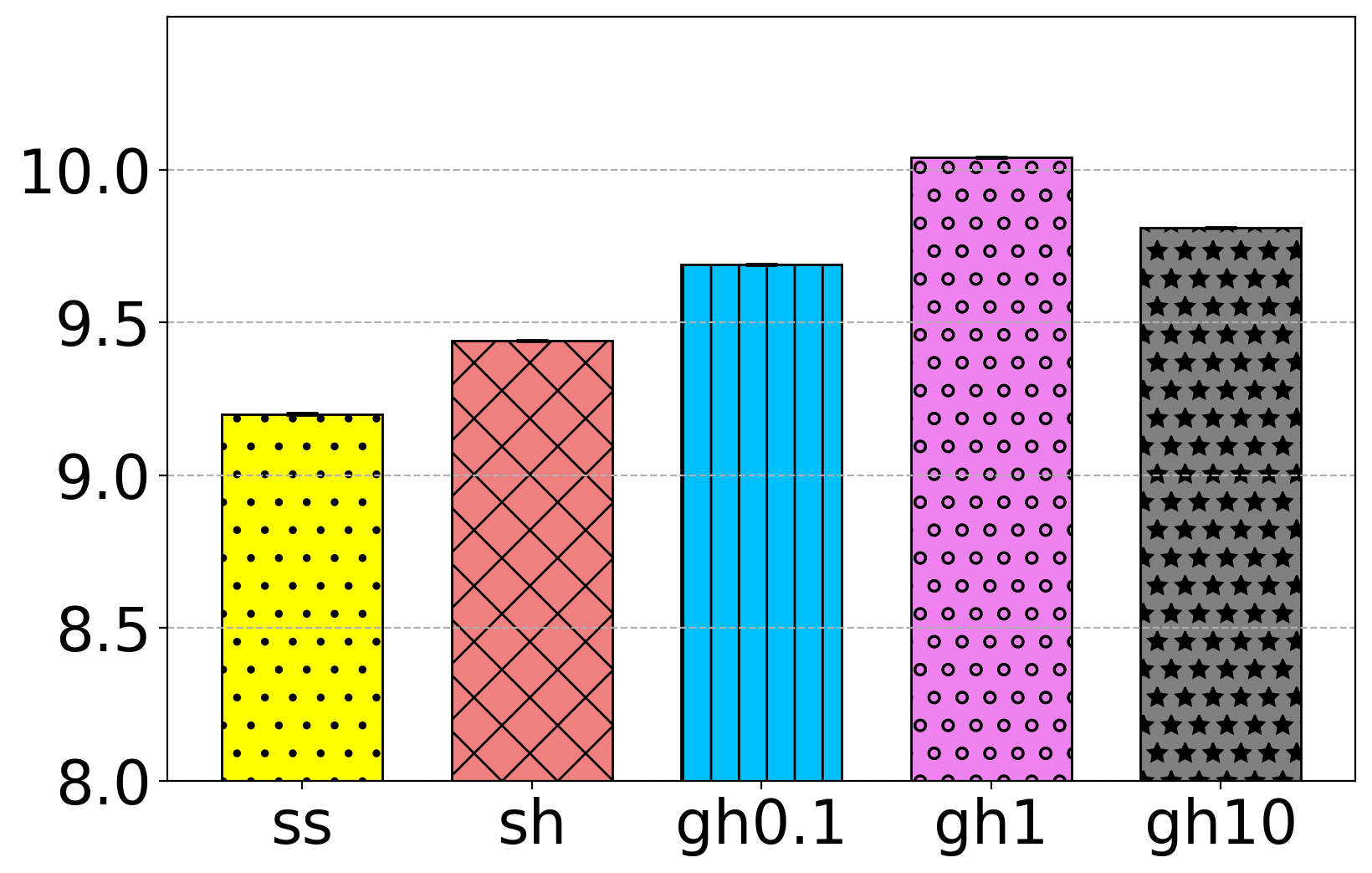}}\\
        \hspace{10pt}\subfloat[Robustness(noise)]
    {\includegraphics[width=.46\columnwidth]{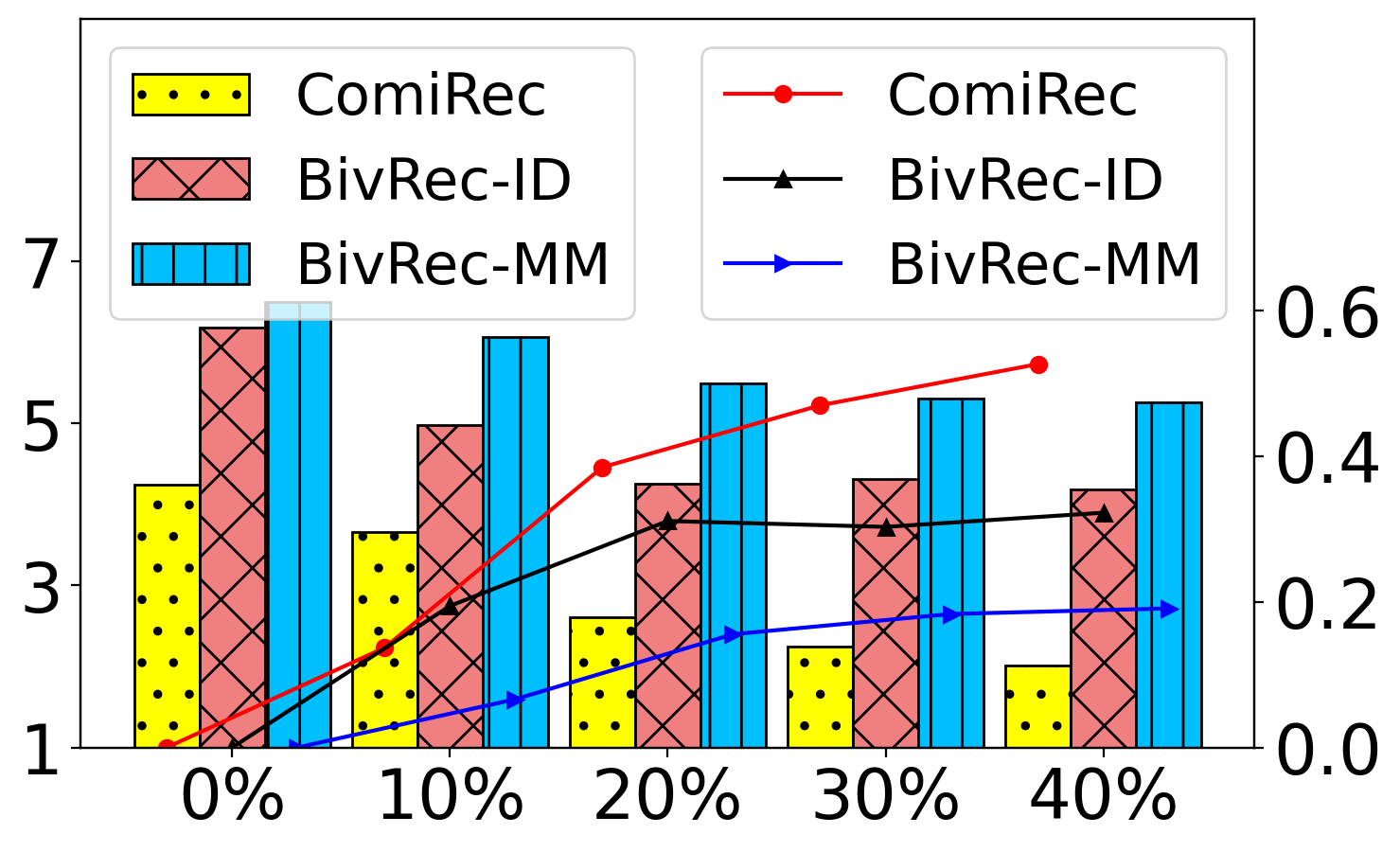}}\hspace{0pt}
        \subfloat[Robustness(cold-start)]
    {\includegraphics[width=.47\columnwidth]{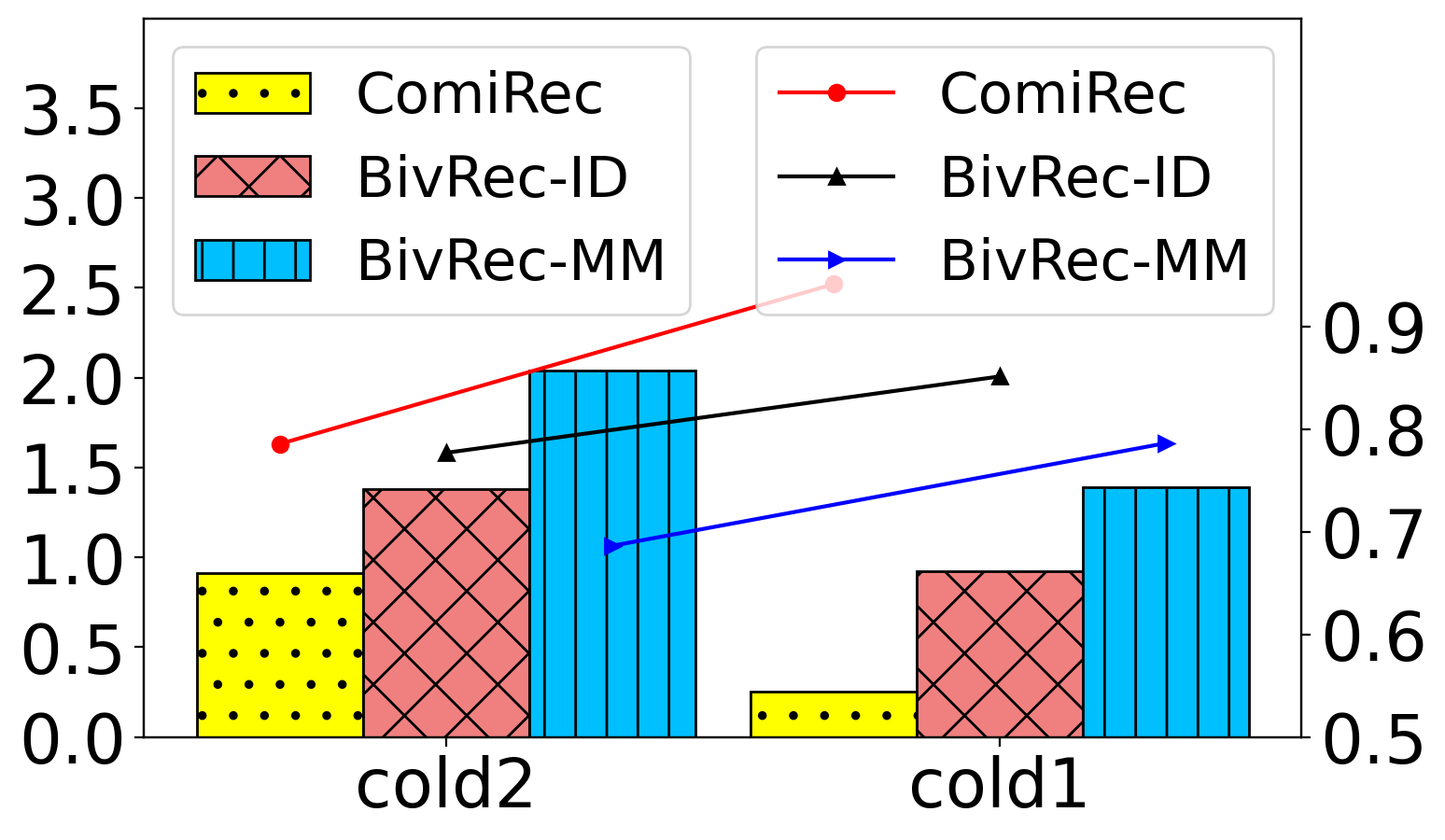}}
     \vspace{-3mm}
	\caption{Experiments results corresponding to Section \ref{3.3.2}, \ref{3.3.3}, and \ref{3.4} respectively. the metric is NDCG@20. (ab) are conducted on the Ml-1m dataset, (cd) are on the Baby dataset.}
        \label{experiments}
        \vspace{-5mm}
\end{figure}

\subsection{Ablation Study (RQ3)}
To answer \textbf{RQ3}, we conduct module ablation experiments, modality ablation experiments, and cluster attention configuration experiments. Additionally, we visualize the interest representations.

\noindent{\textbf{Module Ablation:}}
We systematically remove or replace each component of BivRec with the corresponding component from ComiRec to assess the contribution of each module. As shown in Table \ref{table:each block}: (i) Each module positively impacts the enhancement of recommendation performance. (ii) Notably, the Cluster Attention module exhibits the most substantial influence, resulting in a maximum performance decline of 24.64\% when omitted. Furthermore, this impact is magnified when simultaneously training two recommendation tasks. (iii) When exclusively training with a single type of information, these modules assume a more pronounced role due to the enriched user interests inherent in multimodal data. (iv) It becomes evident that relying solely on contrastive learning to acquire overall semantic similarity is inadequate. Thus, an additional $\mathcal{L}_{A}$ is necessary to learn the process of interest information allocation.

\noindent{\textbf{Modality Abalation:}}
\label{3.3.2}
We investigate the impact of different modalities on the recommendation performance of BivRec. As shown in Figure \ref{experiments} (a), we find out that for BivRec-ID, using both image and text modalities can learn more comprehensive synergistic relationship and achieve the best performance improvement. However, for BivRec-MM, using a single modality is more effective and text information is more suitable for mining user interest than image information. This result is somewhat different from the results reported in previous work, which may be due to the negative impact of simple concatenation operations.

\noindent{\textbf{Temperature Selection for Interest Allocation:}}
\label{3.3.3}We conduct an investigation into the impact of different interest allocation modes in the Cluster Attention. Since the impact on BivRec-ID and BivRec-MM are similar, we only present the results for BivRec-ID here. As shown in Figure \ref{experiments} (b), we test different configurations including softmax with soft gradient(ss), softmax with hard gradient(sh), and Gumbel softmax with a hard gradient(gh) using different values of temperature $\boldsymbol{\tau}$, specifically $\boldsymbol{\tau}$ = 0.1, 1, and 10, respectively. Our experimental results indicate that the proposed Cluster Attention significantly improves the model's performance. The sh model performs better than the ss model, suggesting that controlling the number of interests per item belongs to is effective. BivRec gains exploratory behavior when adding Gumbel softmax noise, leading to further performance improvements. Moreover, by controlling the temperature coefficient, we can adjust the interest allocation process, and as the temperature coefficient increases, the model's exploratory behavior also increases. However, if the temperature coefficient becomes too large, the exploratory behavior can lead to uniform interest allocation, negatively affecting the model's performance.

\noindent{\textbf{Visualization:}}
\begin{figure}[t]
	\centering
	\subfloat[ComiRec-SA]{\includegraphics[width=.415\columnwidth]{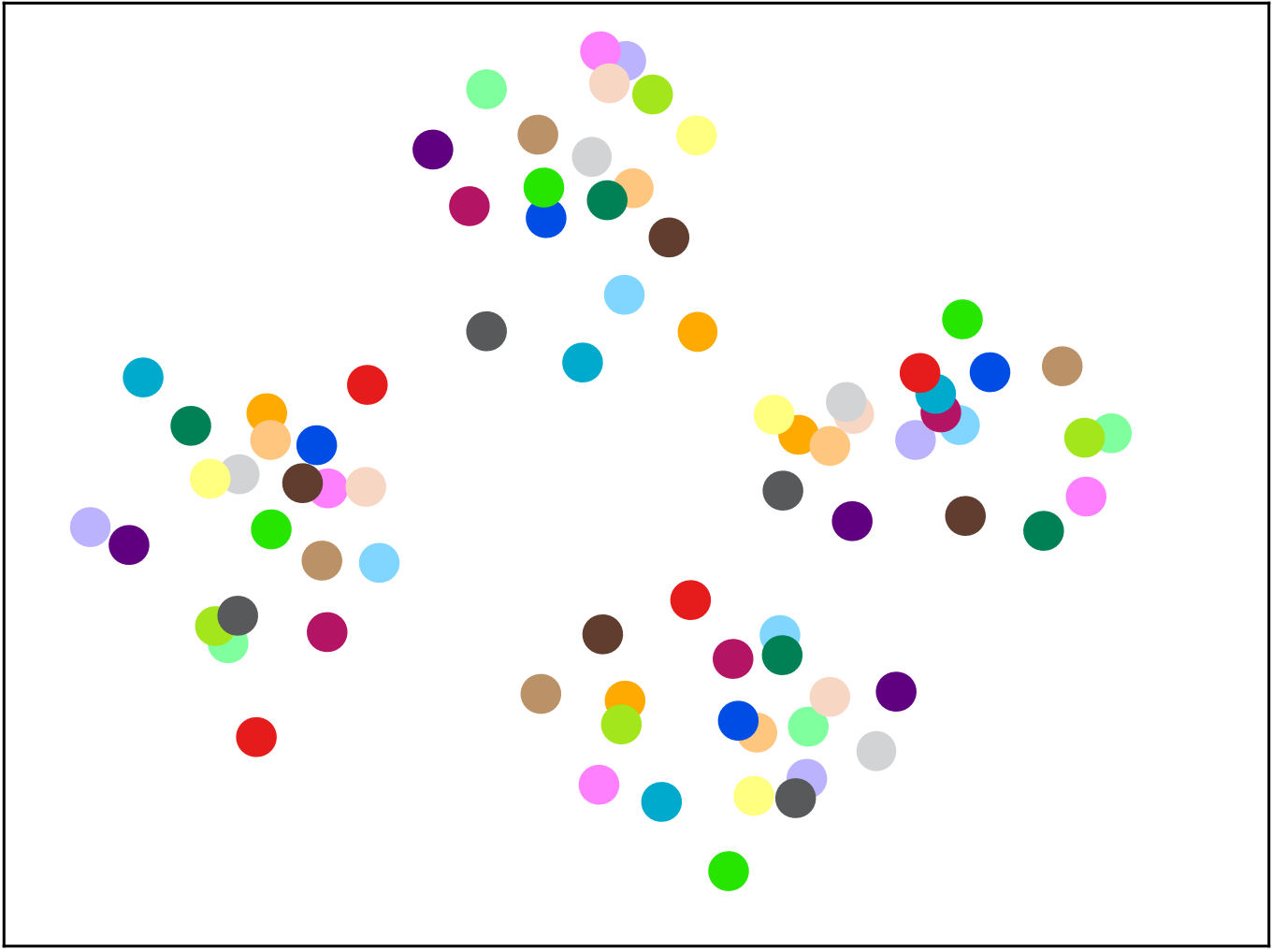}}\hspace{2pt}
	\subfloat[BivRec]{\includegraphics[width=.415\columnwidth]{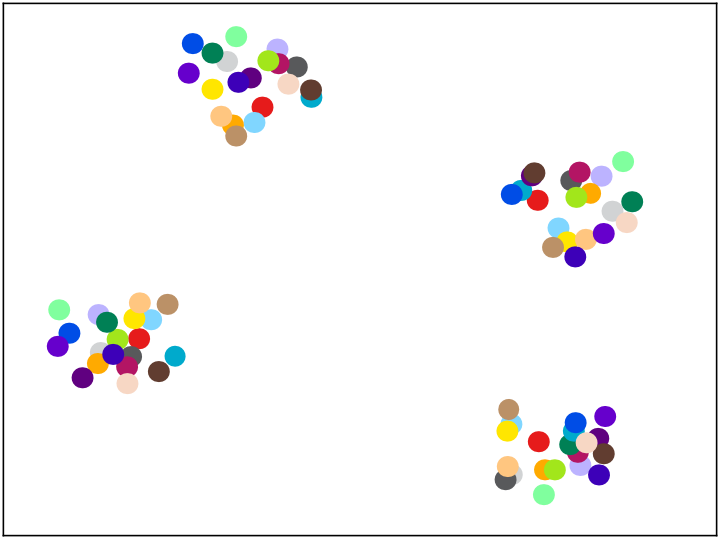}}\\
         \vspace{-3mm}
	\caption{Interest visualization on Ml-1m dataset by T-SNE.}
        \label{fig_tsne}
        \vspace{-5mm}
\end{figure}
To visually validate the extent to which the VID module has generated highly structured interest representations, we have chosen the same 20 users from the ML-1m dataset to visualize their interest representations and compare the visualization results between the BivRec-ID and ComiRec model.
As depicted in Figure \ref{fig_tsne}, each data point corresponds to an interest representation, with interests from the same user visualized using the same color. Upon examination, it becomes apparent that the interests constructed by ComiRec's multi-head attention mechanism do not necessarily capture distinct aspects of interests. Many interests are closely located to one another in terms of distance (e.g., the orange user), resulting in a disorganized representation. This lack of structure poses a challenge in effectively learning and leveraging synergistic relationships. In contrast, the interest representations generated by BivRec-ID exhibit a remarkable level of structure. This structured representation addresses the inherent difficulty of learning synergistic relationships from heterogeneous information, leading to improved performance. 

\subsection{Further Analysis (RQ4)}
\label{3.4}
\begin{table}[t]
        \centering
	\caption{The performance of BivRec-MM in cross-dataset recommendation task, metrics are Recall@20 \%.}
        \vspace{-1em}
        \setlength{\tabcolsep}{10pt}
	\label{zero}
        \resizebox{1\columnwidth}{!}{
        \begin{threeparttable}
	\begin{tabular}{ccccc}
		\toprule[1.5pt]
		\textbf{{Pre-trained model}} & \textbf{{Elec}}  & \textbf{{Cloth}} & \textbf{{Baby}}\\ 
        \midrule
        \midrule
        \ {Supervised Rec} &5.18 &2.13 &4.69 \\
        \midrule
        \ {Ml-25m-ComiRec-SA+} &0.52 &0.37 &0.42 \\
        \ {Ml-25m-image} &0.98 & 0.96  & 0.87  \\
  	\ {Ml-25m-text} &1.18 &0.82 &0.98 \\
        \ {Ml-25m-both} &\textbf{1.24} &\textbf{1.09} &\textbf{1.02} \\
        \ {Improve} &138.46\% &194.59\% &142.86\% \\
        \bottomrule[1.5pt]
	\end{tabular}       
\end{threeparttable}
}
\vspace{-1em}
\end{table}
To answer \textbf{RQ4}, we perform a robust analysis and a cross-dataset recommendation task test for BivRec, aiming to validate whether BivRec preserves the distinct recommendation characteristics inherent in multimodal information. As shown in Figure \ref{experiments} (c) and (d), the left y-axis represents the recommendation metrics, while the right y-axis represents the percentage of performance degradation.
\noindent\textbf{Robustness \textit{w.r.t.} noise information:}
We evaluate the robustness of BivRec against information interference. Specifically, varying proportions of noisy data are added to the Baby dataset during model training and compare the resulting performance with different noise levels. As shown in Figure \ref{experiments} (c), BivRec-ID exhibits significantly enhanced noise robustness due to its joint training approach. Furthermore, BivRec-MM performs even better in this aspect. When the noise level reaches 20\%, BivRec demonstrates the ability to adapt to the noise, whereas ComiRec experiences a continuous decline in performance with increasing noise levels.

\noindent\textbf{Robustness \textit{w.r.t.} user interaction frequency:}
The cold-start problem remains a significant challenge in recommender systems, particularly for new users or those with limited interaction records. To evaluate the performance of BivRec in addressing this problem, we retain only one or two interaction records of users for testing. As shown in Figure \ref{experiments} (d), both BivRec-ID and BivRec-MM demonstrate excellent performance in cold recommendation scenarios. Remarkably, BivRec's performance in the cold1 scenario surpasses even ComiRec's performance in the cold2 scenario. Besides, BivRec-MM similarly shows higher adaptability than BivRec-MM.

\noindent\textbf{Cross-dataset Recommendation:}
To investigate whether BivRec-MM can apply fixed patterns learned from the multimodal information to other datasets, we pretrain it on the Ml-25m dataset and conduct experiments on three datasets: \textit{Baby}, \textit{Cloth}, and \textit{Elec} to assess the performance of its cross-dataset recommendation capability. We exclude the Ml-1m dataset due to its high similarity with ml-25m and use ComiRec+, which fuses the multimodal information to ID information, as the baseline model for comparison. Table \ref{zero} illustrates that BivRec-MM is effectively learning the information patterns that exist across datasets and is outperforming the approach of fusing multimodal and ID information. This indicates that in the BivRec framework, the multimodal recommendation branch still inherits the transferability of multimodal information. It possesses the potential to achieve cross-dataset recommendations and become a versatile recommendation model.

\subsection{Hyper Parameter Analysis}
\begin{figure}[t]
	\centering
	\subfloat[performance with \textit{K,F}]{\includegraphics[width=.50\columnwidth]{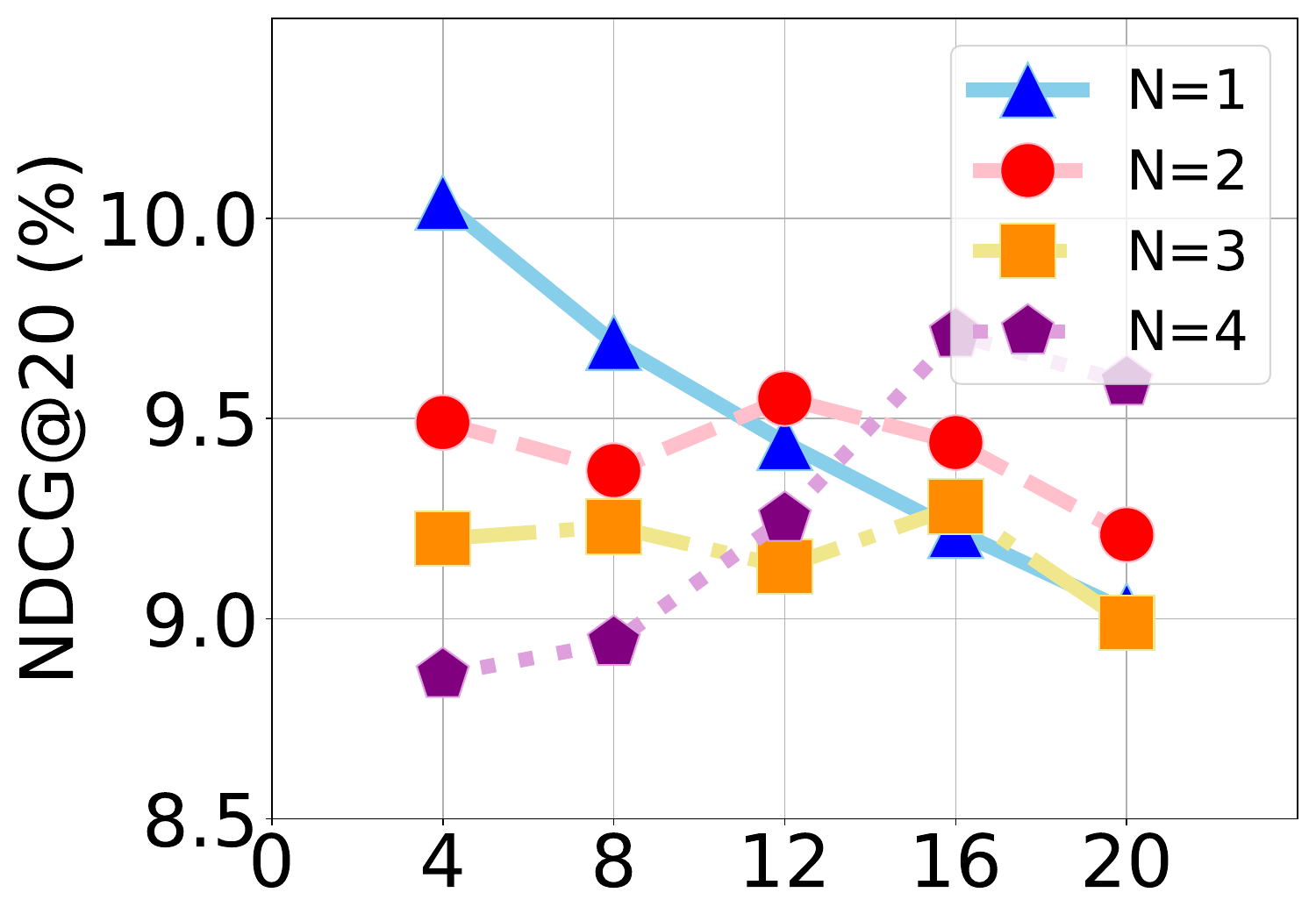}}\hspace{2pt}
	\subfloat[performance with $\lambda_2$,$\lambda_2$]{\includegraphics[width=.47\columnwidth]{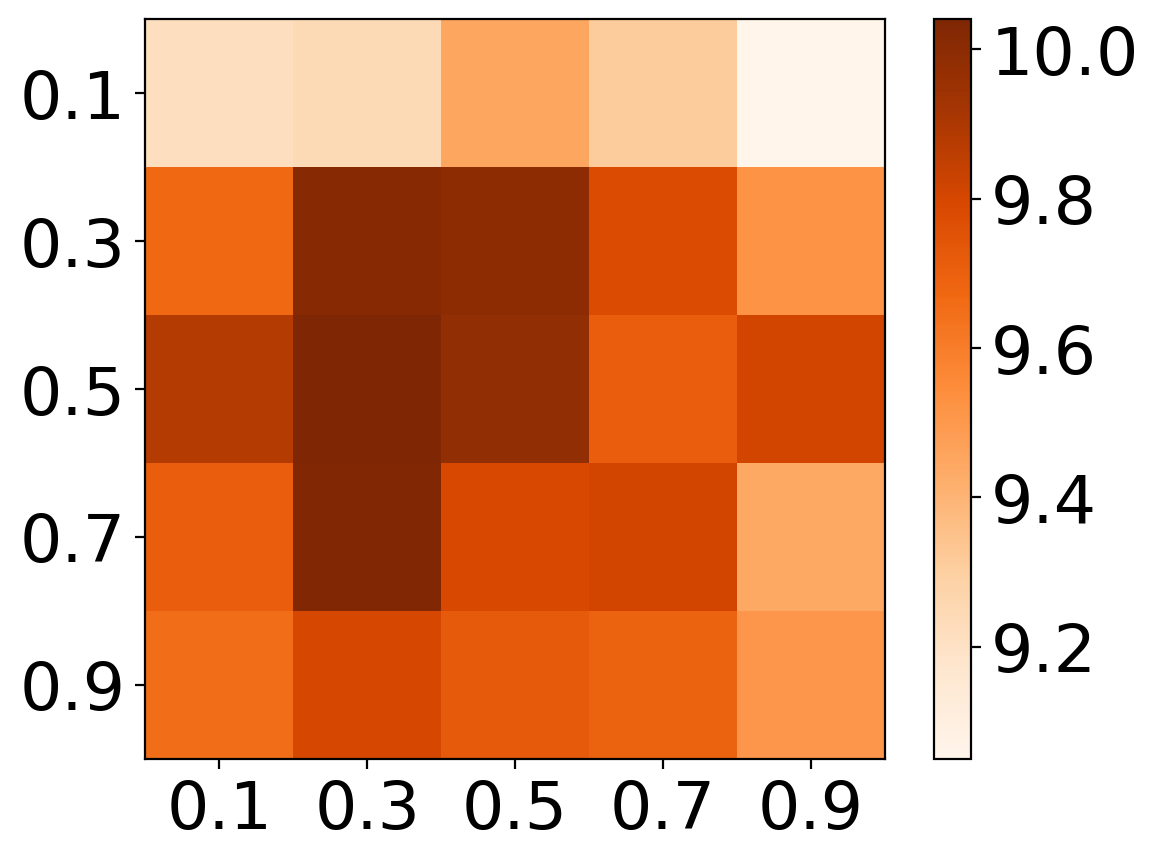}}\\
 \vspace{-3mm}
	\caption{Hyper Parameter Analysis on Ml-1m dataset.}
        \label{hyper}
        \vspace{-4mm}
\end{figure}

We analyze the hyperparameters $K, F, \lambda_1, \lambda_2$ on the Ml-1m dataset. (i) As depicted in Figure \ref{hyper} (a), the hyperparameters ${K, F}$ play a critical role in determining the formation of interest representation by adjusting the number of interests per user and the number of interests each item can belong to. Generally, optimal performance is achieved when the user has four interests, and each item strictly belongs to one interest. Besides, increasing the number of interests $K$ requires an increase in the value of $F$ to ensure effective utilization of interest information. (ii) To analyze the effects of $\lambda_1$ and $\lambda_2$ on the model's performance, we employ a heatmap visualization. As depicted in Figure \ref{hyper} (b), we observe that both extremely large and small values lead to a decline in performance for both parameters. Notably, the heatmap demonstrates a peak that is shifted towards the left region, suggesting that during training, the value of $\lambda_1$ should be slightly smaller than that of $\lambda_2$.

\section{related work}
This section briefly reviews the works related to us,
including sequential recommendation and multimodal recommendation. 

\noindent\textbf{Sequential Recommendations:}
Sequential recommendation tasks (SR) can be classified \cite{tan2021sparse} as single-embedding \cite{kang2018self, sun2019bert4rec, wu2022personalized, chen2022intent, zhang2019feature, xie2022contrastive, zhou2020s3} and multi-embedding models \cite{cen2020controllable, sabour2017dynamic, tan2021sparse, zhang2022re4, shen2021temporal}, depending on the type of user embedding representation used. Early single-embedding SR models employed the Markov chain assumption \cite{he2016fusing, kabbur2013fism, rendle2010factorization, zimdars2013using} to model item transmission relationships and predict the next item based on the last item in the user interaction sequence. Subsequently, GRU4Rec \cite{hidasi2015session} introduced RNN \cite{zaremba2014recurrent} structures to model user temporal interests. With the widespread use of Transformer structures, SASRec \cite{kang2018self} incorporate them into SR tasks, and Transformer-based models have become the mainstream trend. Many models have enhanced SASRec by incorporating contrastive learning \cite{qiu2022contrastive, xie2022contrastive, chen2022intent}, prompt \cite{wu2022personalized, wu2022selective}, and other methods. Although multi-embedding models are initially used in industrial recall layers \cite{cen2020controllable, tan2021sparse, sabour2017dynamic}, using multiple interest expressions of the user for the recommendation task is more interpretable, and recent studies \cite{zhang2022re4, shen2023temporal, li2021intention} have focused on this type of SR task. However, integrating multimodal information into SR tasks remains challenging, and only a few SR models \cite{zhang2019feature, liu2021noninvasive, rashed2022carca} consider multimodal information. Our model effectively addresses this issue by flexibly connecting multimodal and ID sequence information.

\noindent\textbf{Multimodal Recommendations:}
Some existing multimodal sequence recommendation models aggregate multimodal information into ID information as side information. FDSA \cite{zhang2019feature} adaptively combines the two types of information using attention mechanisms. MMMLP \cite{liang2023mmmlp} uses a pure Multi-Layer Perceptron framework. NOVA \cite{liu2021noninvasive} retains the ID information branch separately after fusion to address the issue of information structure invasion. Based on NOVA, DIF-SR \cite{xie2022decoupled} employs a decoupled fusion approach. The multimodal integrating methods have been used in Click-Through Rate prediction tasks like graph pruning \cite{zhou2022tale, liu2023megcf}, fine-grained fusion \cite{lin2019explainable, liu2019nrpa}, multi-view contrastive learning \cite{mu2022learning, ma2022crosscbr} are challenging to be applied in sequential recommendation tasks. Additionally, to overcome the limitations of the first approach's cross-dataset recommendation capability, some models explore pure multimodal recommendation frameworks \cite{yuan2023go}. Models like TransRec \cite{wang2022transrec}, MISSRec \cite{wang2023missrec}, and MMSRec \cite{song2023self} propose multimodal pre-training frameworks. However, these frameworks are still in the early stages of exploration and face challenges of low information utilization and high training costs. In contrast, our BivRec model serves as a new framework that effectively tackles the above issues.
\vspace{-2mm}
\section{concusion}
To tackle the limitations of two types of existing multimodal sequential recommendation frameworks, this paper introduces BivRec, a novel model that considers ID and multimodal information as two distinct views of user interest representation. By jointly training recommendation tasks under both views, BivRec learns the synergistic relationships between interest information in each view, improving the effectiveness of recommendation tasks in both directions. Moreover, BivRec offers flexibility in selecting the information to be used for the final recommendation task. Experimental results demonstrate that BivRec outperforms baseline models on five datasets and performs well in the cold-start recommendation, noise-resistance, and cross-dataset recommendation tasks.
 
\bibliographystyle{ACM-Reference-Format}
\bibliography{reference}
\appendix
\twocolumn
\section{Experimental Setting}
\label{setting}
\subsection{\textbf{Dataset}} We conduct our experiments on five public datasets and we summarize the statistics of the datasets in Table \ref{table:statistics}. The \textit{Ml-1m} and \textit{Ml-25m} datasets \footnote{https://grouplens.org/datasets/movielens/} consist of users' viewing records with movies over time. The \textit{Baby, Cloth, Elec} datasets are subsets of the Amazon Review Dataset \footnote{https://jmcauley.ucsd.edu/data/amazon/}, which is organized by item category. Only users and items with more than five interactions are kept for each dataset \cite{kang2018self}. The image feature is obtained by using Resnet50 \cite{he2016deep}, and the text feature is obtained using Bert \cite{devlin2018bert}.

For data splitting, we adopt the \textit{8:1:1} ratio recommended by multi-embedding sequential recommendation models \cite{cen2020controllable}, where the dataset is divided into three parts for training, validation, and testing respectively. We also follow the data processing method used in ComiRec \cite{cen2020controllable}, where an item is randomly selected from the user sequence data as the target item, and the sequence before this item is used for training. For evaluation, the last 20\% of items in the user sequence are selected.

\subsection{\textbf{Evaluation Metrics}}
We use three metrics commonly used in the recommender system to evaluate the model effect: \textbf{\textit{Recall@k}} measures how many positive cases in the sample are predicted correctly. \textbf{\textit{Normalized Discounted Cumulative Gain@k (NDCG@k)}} measures the ranking of recommendation list. \textbf{\textit{Hit Ratio@k (HR@k)}} measures the percentage that recommend items contain at least one correct item interacted by the user. The \textit{k} is 20. 

\vspace{-1mm}

\subsection{\textbf{Baseline}}
We compare BivRec with general and multimodal sequential recommendation models. To ensure fairness, we modify the sampling and dataset partitioning of the single-embedding model and mark it with a sign *. Besides, we modify the general sequential recommendation model by concatenating \cite{rashed2022carca} the multimodal information onto the ID information with a sign +.

\begin{itemize}[leftmargin=*]
    \item \textbf{ComiRec-SA} \cite{cen2020controllable}: ComiRec-SA adaptively builds multi-interest representations using multi-head attention mechanisms \cite{vaswani2017attention}.
    \item \textbf{SASRec} \cite{kang2018self}: SASRec uses the attention mechanism to explore users' interests at different stages.
    \item \textbf{FDSA} \cite{zhang2019feature}: FDSA designs a feature-granularity attention network. We modify side information to multimodal information.
    \item \textbf{MMMLP} \cite{liang2023mmmlp}: MMMLP employs a pure MLP architecture to facilitate the interaction between multimodal and ID information, utilizing an additive fusion approach for integration. 
    \item \textbf{MMSRec} \cite{song2023self}: A self-supervised learning framework that leverages pure multimodal information for recommendations, we eliminate the pre-training phase and directly employ a two-layer Transformer as the item encoder. 
    \item \textbf{NOVA} \cite{liu2021noninvasive}: NOVA retains the item sequence information separately and integrates the edge information in another network structure for the sake of information intrusion. We modify side information to multimodal information.
    \item \textbf{DIF-SR} \cite{xie2022decoupled}: DIF-SR proposes Decoupled Side-information Fusion Block based on NOVA. We modify side information to multimodal information.
    \vspace{-2mm}
\end{itemize} 

\begin{table}[t]
        \centering
	\caption{Summary of the datasets.}
        \vspace{-2mm}
        \setlength{\tabcolsep}{12pt}
	\label{table:statistics}
        \resizebox{1\columnwidth}{!}{
        \begin{threeparttable}
	\begin{tabular}{ccccc}
		\toprule[1.5pt]
		\textbf{Data} & \textbf{\#Users} & \textbf{\#Items} & \textbf{\#Avg-length} & \textbf{\#Interactions}\\ 
        \midrule
        \midrule
		\ Ml-25m  & 162,541 &   62,423  & 154  &25,000,095 \\
        \ Ml-1m &6,040 & 3,417  & 165 &1,000,209 \\
  	\ Elec &192,402 &63,000 &8 & 1,689,188\\
        \ Cloth &39,386 &23,032 &7 & 278,677\\
        \ Baby &19,444 &7,049 &8 &160,792 \\
        \midrule
        \ Book &459,133 &313,966 &19 &8,898,041 \\
        \ Taobao &976,779 &1,708,530 &87 &85,384,110 \\

        \bottomrule[1.5pt]
	\end{tabular}       
\end{threeparttable}
}
\vspace{-2mm}
\end{table}

\subsection{\textbf{Implementation Details}}
All models are implemented in PyTorch and optimized using the Adam optimizer in NVIDIA RTX3090-24G and A40-48G. The maximum number of iterations is set to one million. $\lambda$1, $\lambda$2, $F$ are set to 0.5, 0.3, 1. For ML datasets, scale and $K$ are set to [4,8] and [24, 4]. For other datasets, scale and $K$ are set to [4] and [4]. The batch size is set to 2048. The temperature parameters $\beta$, $\phi$ for contrastive learning are 0.07, the $\tau$ is 1, the learning rate is 0.001, and the hidden dimension for both views is 256. Missing multimodal features in the original dataset are filled with zero vectors.

\section{\textbf{Application on Recall Stage}}
\label{recall stage}
In practical applications, multimodal features are often utilized during the sorting stage to provide users with more accurate recommendations. BivRec is also considered to be more similar to a sorting model. However, before that, there exists a recall stage. The initial purpose of the multi-interest recommendation model was to swiftly recommend a set of potential items to users from a vast pool of data in the recall stage. Although most current research does not explicitly differentiate between these two types of models, we will explore the application of BivRec in the recall phase. Since the primary focus of the recall phase is efficiency, we simplify BivRec by retaining only the ID information and Cluster Attention. We will compare its performance with the mainstream baseline model ComiRec using two widely used industry-scale datasets: Amazon Book and Taobao dataset.

From Table \ref{recall}, it is evident that the Cluster Attention module is highly suitable for the recall task. It exhibits a significant improvement in performance compared to the ComiRec-SA baseline. Additionally, during actual training, it reduces the training time by more than 20\%.

\begin{table}[h]
\centering
\caption{The performance in the Recall Task.}
\vspace{-2mm}
\setlength{\tabcolsep}{4pt}
\resizebox{0.95\columnwidth}{!}{
\begin{tabular}{c|ccc|ccc}
\toprule[1.5pt]
\multicolumn{1}{c}{\textbf{Model}}&\multicolumn{6}{c}
{\textbf{Dataset}}\\
\cmidrule(l){1-1}\cmidrule(l){2-7}
\textbf{Metrics@20}&\multicolumn{3}{c}{Book}&\multicolumn{3}{c}{Taobao}\\
\cmidrule(l){2-4}\cmidrule(l){5-7}
\textbf{Units \% } &Recall & \ \ \ NDCG &HR &Recall & \ \ \ NDCG &HR\\
\midrule
\midrule
\multirow{1}{*}{ComiRec-SA} 
&5.49 &8.99 &11.40 &6.90 & 24.68& 41.55\\
\midrule
 \multirow{1}{*}{$PureID_\text{Cluster Attention}$} 
&7.62 & 10.59& 14.47& 8.31& 28.20 & 46.73 \\
\bottomrule[1.5pt]
\end{tabular}}
\label{recall}
\end{table}

\section{Case Study}
\label{case}
\begin{figure}[t]
    \centering
    \includegraphics[width=0.9\linewidth]{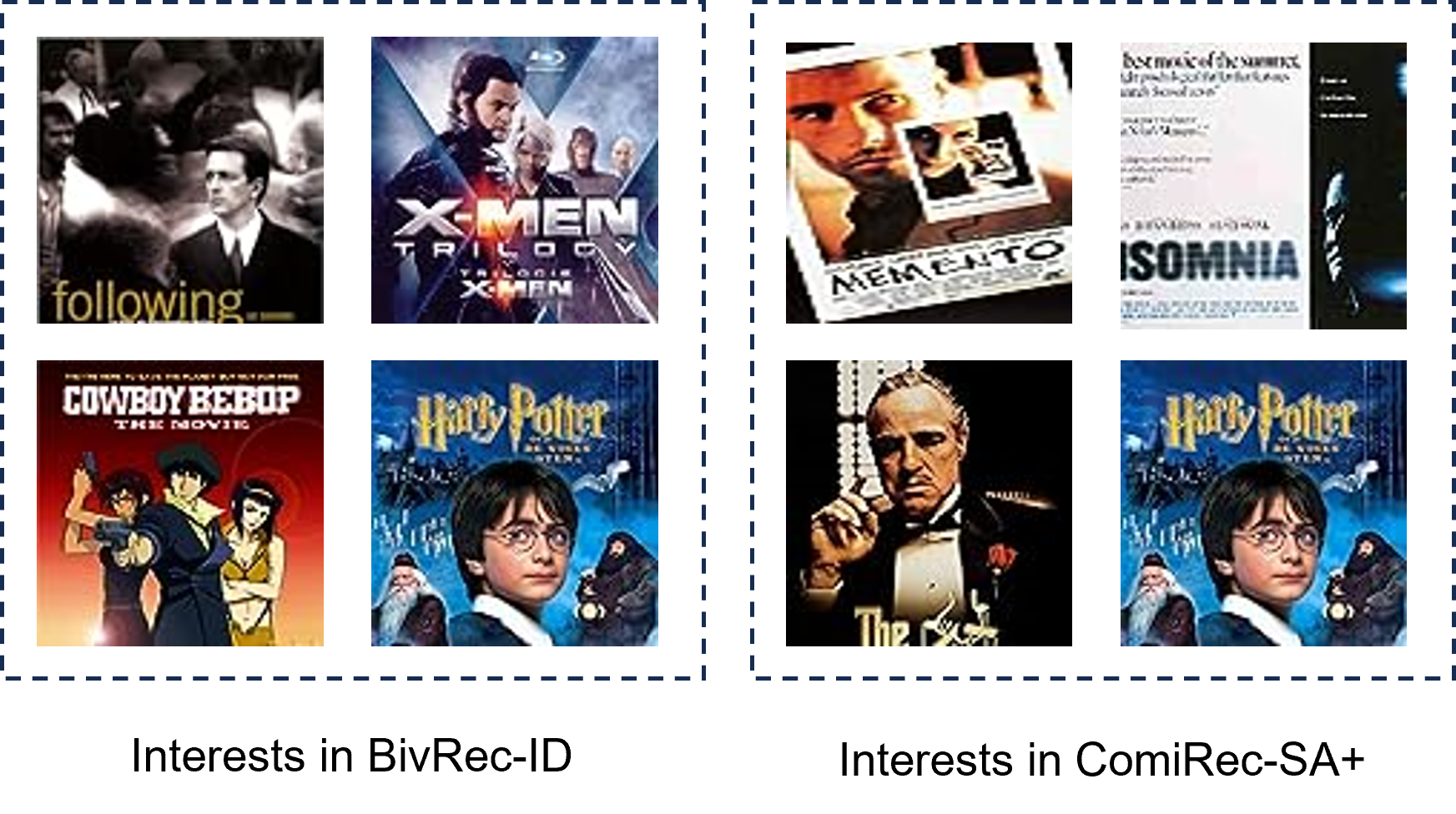}
    \vspace{-3mm}
    \caption{Case Study in ML-1m dataset.} 
    \label{fig:case}
    \vspace{-3mm}
\end{figure}
\vspace{-1mm}
Based on Figure \ref{fig:case}, we visualize user interests by determining the closest movies representation using BivRec-ID and ComiRec-SA+ on the ML-1m dataset. It is evident that while both methods incorporate multimodal information, BivRec-ID demonstrates a more effective utilization of this information. The interest representations generated by BivRec-ID offer a clearer and more intuitive understanding of user interests. Conversely, the interests constructed by ComiRec-SA+ may appear somewhat ambiguous.

\section{Algorithm}
\begin{algorithm}[h]
	\caption{\label{Algorithm} Optimization Algorithm of BivRec.}
	\raggedright
	{\bf Input}: ID sequence $\{\boldsymbol{S}_{u}\}_{u=1}^{ \mathcal{U}}$, MM sequence $\{\boldsymbol{S}_{u}^{m}\}_{u=1}^{ \mathcal{U}}$, hyper-parameters $K,F,\lambda$ and scales $\{S\}$.\\
	{\bf Output}: model parameter $\theta\{\theta_{ID}, \theta_{MM} \}$for BivRec-ID and BivRec-MM\\
	\begin{algorithmic} [1]
        \STATE randomly initializes $\theta$ and set Iteration = 0
		\WHILE{Iteration $\leq$ MaxIteration}
            \STATE Get $\boldsymbol{E}^{id}$, $\boldsymbol{E}^{m}$ via Multi-scale Interest Embedding with $\{S\}$.
            \STATE Get $\boldsymbol{V}$, $\boldsymbol{V}_{m}$ via Eq. \ref{get interest} to compute $\mathcal{L}_{Rec}^{ID}$ and $\mathcal{L}_{Rec}^{MM}$.
            \STATE Get $\boldsymbol{A}_{A}^{id}$, $\boldsymbol{A}_{A}^{m}$ via Eq. \ref{Allocation} to compute $\mathcal{L}_{A}$.
            \STATE Get $\hat{\boldsymbol{h}}$, $\hat{\boldsymbol{h}}_m$ via Eq. \ref{unit} to compute $\mathcal{L}_{MM\leftrightarrow ID}$.
            \STATE Optimize the model via Eq. \ref{loss}.
            \STATE Iteration +=1
		\ENDWHILE
	\end{algorithmic}
\end{algorithm}

\end{document}